\documentclass[12pt]{amsart}

\usepackage{amssymb,amsmath,amscd,color}

\theoremstyle{plain}

\theoremstyle{definition}

\theoremstyle{remark}

\newtheorem*{remark*}{Remark}

\newcommand{\cU}{{\mathcal{U}}}
\newcommand{\cV}{{\mathcal{V}}}

        \newcommand{\field}[1]{{\mathbb{#1}}}
        \newcommand{\NN}{\field{N}}
        \newcommand{\ZZ}{\field{Z}}
        
        \newcommand{\RR}{\field{R}}
        \newcommand{\CC}{\field{C}}

\allowdisplaybreaks

\begin{document}

\title[Quasiclassical approximation for magnetic monopoles]{Quasiclassical approximation for magnetic monopoles}

\author[Y. A. Kordyukov]{Yuri A. Kordyukov}
\address{Institute of Mathematics, Ufa Federal Research Centre, Russian Academy of Sciences, 112~Chernyshevsky str., 450008 Ufa, Russia and Novosibirsk State University, Pirogova st 1, 630090, Novosibirsk, Russia} \email{yurikor@matem.anrb.ru}

\author[I. A. Taimanov]{Iskander A. Taimanov}
\address{Sobolev Institute of Mathematics, 4 Acad. Koptyug avenue, and Novosibirsk State University, Pirogova st 1, 630090, Novosibirsk, Russia}
\email{taimanov@math.nsc.ru}

\thanks{This work was supported by the Laboratory of Topology and Dynamics, Novosibirsk State University (contract no. 14.Y26.31.0025 with the Ministry of Education and Science of the Russian Federation).}



\begin{abstract}
A quasiclassical approximation is constructed to describe the eigenvalues of the magnetic Laplacian on a compact Riemannian manifold in the case when the magnetic field is not given by an exact $2$-form. For this, the multidimensional WKB method in the form of Maslov canonical operator is applied. In this case, the canonical operator takes values in sections of a nontrivial line bundle. The constructed approximation is demonstrated for the Dirac magnetic monopole on the two-dimensional sphere.
\end{abstract}

\dedicatory{To V.V. Kozlov on his 70th birthday}

\date{}

 \maketitle
\section{Introduction}

Magnetic monopoles were introduced by Dirac at the beginning of the 1930s in \cite{Dirac}. In particular, it was said there (see \cite[p. 71]{Dirac}) that:
\smallskip\par
{\sl ``Elementary classical theory allows us to formulate equations of motion for
an electron in the field produced by an arbitrary distribution of electric charges
and magnetic poles...

The object of the present paper is to show that quantum mechanics does not
really preclude the existence of isolated magnetic poles.
On the contrary, the present formalism of quantum mechanics ... , when developed naturally
without the im\-po\-si\-tion of arbitrary restrictions
leads inevitably to wave equations whose only physical interpretation is the motion of an electron in the
field of a single pole. This new development requires no change whatever in
the formalism ... Under these cir\-cum\-stan\-ces one would be surprised if Nature had made no use of it.

The theory leads to a connection ... between the quantum
of magnetic pole and the electronic charge.''}
\smallskip\par
Dirac considered the motion of the electron in the magnetic field
$$
{\bf H} = \frac{\mu {\bf r}}{r^3}, \ \ \ \mu = \mathrm{const},
$$
which is directed radially and has singularity at the origin (for $r=0$). He showed that systems, which include such magnetic fields (corresponding, probably, to several singularities) are quantizable only if the flow of the magnetic field through a small sphere around the singularity (magnetic monopole) satisfies the condition
\begin{equation}
\label{diracN}
4\pi \mu = 2\pi N \frac{\hbar c}{e},
\end{equation}
where $N$ is some integer, $\hbar$ is the Planck constant, $c$ is the speed of light  and $e$ is the charge of electron, that gives a relationship between the magnetic charges of the monopoles and the charge of electron:
$$
e \, \mu  = \mathrm{const} \, N.
$$
It follows from here that
\smallskip\par
{\sl ``... if there exists any monopole at all in the universe, all electric charges would have to be such that $e$ times this monopole
strength is equal to $\frac{1}{2}n\hbar c$'' \cite[p. 240]{Dirac1978}.}
\smallskip\par
The ideas of \cite{Dirac} were developed in the paper by Tamm \cite{Tamm} (note that these two papers quoted each other as papers in print), where the spectrum of the quantum Hamiltonian for one singularity was computed. By separation of variables, the problem is reduced to a one-dimensional (radial) problem and a two-dimensional spectral problem on the sphere (in modern terminology, to the magnetic Laplacian). The eigenfunctions of the magnetic Laplacian found by Tamm have singularities.

The presence of singularities in Tamm's computations was explained in the paper by Wu and Yang \cite{WY}, where it was shown that, in fact, the magnetic Laplacian acts on sections of $U(1)$--bundles $L^N$ over the sphere, and, for it, the magnetic field --- the $2$-form $F$ on the sphere --- is the curvature form. Since the Chern class of such a bundle is integer, this explains from a topological point of view quantization of the magnetic field flow  $N = \frac{1}{2\pi} \int_{S^2}F$ (this formula becomes \eqref{diracN}, if we put, as it is often done, $\hbar = c = e = 1$).

If $N \neq 0$, then the vector potential of the magnetic field $A = d^{-1} F$ is not defined on the whole sphere. But it is defined on the once punctured sphere and, in \cite{Tamm}, the eigenfunctions of the magnetic Laplacian were computed in such a domain $U = S^2 \setminus \{\mathrm{pt}\}$; they are not extended without singularities to the puncture. It shown in \cite{WY} how to get eigenfunctions on the whole sphere from Tamm's computations: it is necessary to remove from it a pair of points $a$ and $b$, corresponding to the values $\theta =0$ and $\theta = \pi$ of the spherical coordinate $\theta$, and take  ``Tamm's eigenfunctions'' $\psi_a$ and $\psi_b$ in $U_a = S^2 \setminus\{a\}$ and $U_b = S^2 \setminus \{b\}$, which are glued together into an ``eigensection'' of the magnetic Laplacian on the sphere. Here, on the intersection $U_a \cap U_b$, the following identity holds:
 $$
 \psi_a = e^{iN\varphi} \psi_b.
 $$

Observe that the existence of magnetic charges in the language of classical dynamics was apparently first considered by Poincar\'e \cite{Poincare}. The most complete survey of the current state of monopole theory was done in \cite{Shnir}.

In this paper, we will consider systems determined by a compact manifold $M$ equipped with a Riemannian metric $g_{jk}$ and a closed $2$-form $F$  (magnetic field). Such a system describes the motion of a charged particle on $M$ in the magnetic field defined by the $2$-form $F$. Its classical dynamics given by the magnetic geodesic flow  --- the Hamilton system on the cotangent bundle $T^\ast M$ of $M$ with Hamiltonian
$$
H(x,p) =\sum_{j,k=1}^n g^{jk}(x) p_j p_k, \quad x \in M, p \in T^\ast_xM,
$$
with respect to the (twisted) symplectic form
$$
\Omega = \sum_{j=1}^n dp_j\wedge dx^j + F.
$$
(cf.  \cite{Novikov1982}).
Following the conventional terminology in mathematical physics, we will talk about the magnetic field as a magnetic monopole in the case when the form $F$ is non-exact.

To quantize such a classical system, it is necessary that the de Rham cohomology class of the form $\frac{1}{2\pi}F$ is integer:
 $$
 \left[\frac{1}{2\pi} F\right] \in H^2(M;\ZZ).
 $$
In this case, it is the first Chern class of a line bundle $L$ on $M$:
 $$
 c_1(L) = \left[\frac{1}{2\pi} F\right],
 $$
and the magnetic Laplacian (the quantum Hamiltonian of the system) acts on sections of the bundle $L^N$ (see, for instance, \cite{KT}).

The interest in periodic trajectories of magnetic geodesic flows, including magnetic mono\-poles, which appeared after the papers \cite{Novikov1982, NS}, is largely due to the fact that such systems are obtained as the reductions of natural mechanical systems, describing the motions of rigid bodies (tops) such as the Kirchhoff problem on the motion of a rigid body in an unbounded ideal fluid \cite{NS}, the problem on the rigid body rotation with a fixed point in an axisymmetric field \cite{K85} and others. A magnetic monopole arises in the case when such a system is restricted to a nonzero level of some first integral $I$ (``area integral '' in the case of the rigid body rotation problem). Here the magnetic geodesic flow on the two-dimensional sphere appears and $\int_{S^2}F=4\pi I$. In particular, when the system is restricted to the zero level of such an integral, the resulting magnetic field is exact and the flow is equivalent to a
classical natural mechanical system (for the rigid body rotation problem, this situation is studied in detail in \cite{K76}).

The main goal of the paper is to construct a quasiclassical approximation for describing the spectrum of the magnetic Laplacian. Formal application of the multidimensional WKB method in the form of Maslov canonical operator \cite{Maslov65, Maslov-Fed} to the magnetic Laplacian leads to the following:

1) Lagrangian manifolds should be considered with respect to the symplectic form
$\Omega $, which requires making corrections in the quantization conditions;

2) the action $S$ becomes multi-valued for non-exact magnetic fields:
for example, for a chart $\cU$ on a Lagrangian manifold $\Lambda $, invariant under the magnetic geodesic flow, which is uniquely projected onto a domain in $M$ under the canonical projection $\pi: T^*M \to M$, the canonical operator takes the form
$$
K^h_{\Lambda,\cU} u(x)=e^{(i/h)S_\cU(s)}\sqrt{\left|\frac{d\mu(s)}{dx_g}\right|}u(s), \quad x\in \pi(\cU),
$$
where $s$ is a unique point in $\cU\subset \Lambda$ such that $\pi(s)=x$, $d\mu$ is a smooth measure on $\Lambda$, invariant under the magnetic geodesic flow, $dx_g$ is the Riemannian volume form,
$$
S_\cU(s) = \int^s_{s_0} d^{-1} \Omega = \int^s_{s_0} \left(\sum_{j=1}^n p_j dx^j + A_{U}\right),\quad s\in \cU,
$$
$A_U$ is a 1-form, satisfying the condition
$$
dA_U = F
$$
in some neighborhood $U$ of $\pi(\cU)$ in $M$
and $h$ is a small parameter. In the quasiclassical approximation $h$ is the Planck constant and for magnetic monopoles $h$ takes a discrete set of values:
$$
h=\frac{1}{N}, \ \ \ \mbox{where $N$ is an integer}, 
$$
because of the quantization condition on the magnetic field.
Multi-valuedness of the integrand in the definition of $K^h_{\Lambda, \cU}$ arises due to non-exactness of the $2$-form $F$: the equation $d^{-1}F = A$ has no solutions, defined globally on the whole manifold $M$. As shown below in \S 2,
this is not an obstruction to constructing the canonical operator:
\smallskip\par
{\sl the fact that $K^{1/N}_{\Lambda, \cU}u$ is not single-valued is related to the fact that the result of applying of this operator to a function on $\Lambda$ will be a section of a nontrivial bundle, and, for example, in the case of two overlapping charts
$\cU$ and $\cV$, the formulas for $K^{1/N}_{\Lambda, \cU}u$ and $K^{1/N}_{\Lambda, \cV}u$ will give two coinciding sections of the same bundle $L^N$ over the intersection $\pi(\cU) \cap \pi(\cV)$, but with respect to different trivializations. Therefore
\begin{equation}
\label{1}
K^{1/N}_\Lambda: C^\infty(\Lambda) \to C^\infty(M,L^N).
\end{equation}
}
\smallskip\par
A detailed construction of the canonical operator (\ref{1}) for the magnetic Laplacians is given in \S 2.

In \S 3 we apply the constructions of \S 2 to the Dirac magnetic monopole on the two-dimensional sphere. In this case, the almost eigenvalues of the magnetic Laplacian constructed by using the quasiclassical approximation are shown to coincide with its exact eigenvalues up to a constant correction term.

In the absence of the magnetic field, the eigenfunctions of the Lap\-la\-ce-Bel\-tra\-mi operator are the spherical polynomials and the problem is exactly solvable.
In \cite{Kogan}, the canonical operator method (for demonstration purposes) was applied to this system and it was shown that it is possible to find all eigenvalues up to a bounded constant with its help as well as to obtain quasiclassical approximations of eigenfunctions.
This approach is based on complete integrability of the geodesic flow on the two-dimensional sphere and, therefore, on the existence of a foliation of almost the whole phase space (except for a measure zero set) into invariant Lagrangian tori.

The geodesic flow on the unit sphere and the magnetic geodesic flows on constant curvature surfaces with constant magnetic field are superintegrable:
all trajectories are closed and are geodesic circles. Therefore, there are different ways to glue together these circles into invariant tori, not necessarily, along the level surfaces of a smooth function on the centers of closed trajectories. The choice of an additional first integral of the flow determines one or another quasiclassical approximation for the system.

Using invariant tori satisfying the quantization conditions, almost eigenfunctions are constructed. These functions are concentrated modulo $O(N^{-\infty})$ on the projections of tori. For a fixed $j$ and $N \to \infty$ the projections of tori corresponding to the almost eigenvalue $\hat{\lambda}_{N, j}$ (that is, the $j$th eigenvalue) concentrate around the curves in which the centers of closed trajectories are projected, but, nevertheless, the union of the projections of all such tori covers the whole manifold, since their number grows linearly as the multiplicity of the eigenvalue.

We remark that a construction of a canonical operator on an arbitrary symplectic manifold was proposed in \cite{Karasev-Maslov}. However such an operator takes values in a certain
bundle of wave packets in difference with (\ref{1}) where the canonical operator takes values in the same space as the approximated operator. In \cite{BNSh}, the operator from \cite{Karasev-Maslov} was used for constructing spectral series for the quantum Hamiltonian of a charged particle in constant nonzero magnetic field on a hyperbolic surface. In this case, the magnetic field becomes exact on the universal covering of the surface, the particle trajectories are geodesic circles, any smooth function on the surface defines a first integral of the flow \cite{T} ---  a closed trajectory corresponds to the value of the function in the center of the geodesic circle, and, starting from any such smooth function, one can construct a quasiclassical approximation to describe the series of eigenvalues of the magnetic Laplacian below some threshold level. For another nontrivial system, namely, for the motion of a charged particle in strong constant magnetic field on the plane in the presence of periodic electric field, spectral series of the quantum problem were constructed in \cite{BDP}.

\section{Quasiclassical approximation for the magnetic Laplacian}\label{s:Maslov}
Let $(M,g)$ be a compact Riemannian manifold of dimension $n$ equipped with a closed $2$-forn $F$. Consider a Hermitian line bundle $(L,h^L)$ on $M$ with a Hermitian connection $\nabla^L : C^\infty(M,L)\to C^\infty(M,T^*M\otimes L)$. The curvature form of $\nabla^L$ is given by $R^L=(\nabla^L)^2$. We will assume that it is related with the magnetic field form $F$ by
\begin{equation}\label{e:prequant}
F = iR^L.
\end{equation}
It is well-known that if the form $F$ satisfies the quantization condition
\begin{equation}\label{e:prequantum}
\left[\frac{1}{2\pi} F\right] \in H^2(M;\ZZ),
\end{equation}
then such a Hermitian line bundle $(L,h^L)$ with Hermitian connection $\nabla^L$ exists.

The Riemannian metric on $M$ and the Hermitian structure on $L$ allow us to define inner products on $C^\infty(M,L)$ and $C^\infty(M,T^*M\otimes L)$ and the adjoint operator $(\nabla^L)^* : C^\infty(M,T^*M\otimes L)\to C^\infty(M,L)$. The magnetic Laplacian is the second order differential operator acting on $C^\infty(M,L)$ by
\[
\Delta^{L} = (\nabla^L)^*\nabla^L.
\]

For any $N\in \NN$, consider the $N$th tensor power $L^N=L^{\otimes N}$ of the line bundle  $L$. Denote by $\Delta^{L^N}$ the corresponding magnetic Laplacian in $C^\infty(M,L^N)$. Let $\{\lambda_{N,j}, j=0,1,2,\ldots\}$ be the non-decreasing sequence of the eigenvalues of  $\Delta^{L^N}$ taken with multiplicities. We are interested in the asymptotic behavior of the eigenvalues $\lambda_{N,j}$ and the corresponding eigenfunctions in the quasiclassical limit  (as $N\to \infty$).  For this, we apply the WKB-method, based on the use of the Maslov canonical operator.

Choose local coordinates $(x^1,\ldots,x^n)$ on an open subset $U$ of $M$. Suppose that the Hermitian line bundle $L$ is trivial on $U$, that is,
\[
L\left|_U\right.\cong U\times \CC\ \text{and}\ |(x,z)|_{h^L}=|z|,\quad (x,z)\in U\times \CC.
\]
Then the covariant derivative $\nabla^L$ can be written as
\[
\nabla^L = d-i A_U : C^\infty(U)\to C^\infty(U,T^*U),
\]
where $A_U= \sum_{j=1}^nA_{U,j}(x)\,dx^j$ is the real-valued connection 1-form  (magnetic potential). It is easy to check that $R^L=-idA_U$ and $F$ is given by
\[
F=dA_U=\sum_{j<k}F_{jk}\,dx^j\wedge dx^k, \quad
F_{jk}=\frac{\partial A_{U,k}}{\partial x^j}-\frac{\partial
A_{U,j}}{\partial x^k}.
\]
Write the Riemannian metric matrix $g$ as $g(x)=(g_{j\ell}(x))_{1\leq j,\ell\leq n}$, its inverse as
$
g(x)^{-1}=(g^{j\ell}(x))_{1\leq j,\ell\leq n}
$
and denote $|g(x)|=\det(g(x))$, then $\Delta^{L^N}$ takes the form
\begin{equation}\label{e:DLp}
\Delta^{L^N} =-\frac{1}{\sqrt{|g(x)|}}\sum_{1\leq j,\ell\leq n}\left(\frac{\partial}{\partial x^j}-iNA_{U,j}(x)\right)\left[\sqrt{|g(x)|}
g^{j\ell}(x) \left(\frac{\partial}{\partial
x^\ell}-iNA_{U,\ell}(x)\right)\right].
\end{equation}

Consider a differential operator $\hat{H}^h_{U}$, depending on a parameter $h>0$:
\begin{equation}\label{e:Hh}
\hat{H}^h_{U} =\frac{1}{\sqrt{|g(x)|}}\sum_{1\leq j,\ell\leq n}\left(\frac{h}{i}\frac{\partial}{\partial x^j}-A_{U,j}(x)\right)\left[\sqrt{|g(x)|}g^{j\ell}(x) \left(\frac{h}{i}\frac{\partial}{\partial x^\ell}-A_{U,\ell}(x)\right)\right].
\end{equation}
The operators $\Delta^{L^N}$ and $\hat{H}^h_{U}$ are related by
\begin{equation}\label{Delta-hatH}
\Delta^{L^N}=h^{-2}\hat{H}^h_{U}, \quad h=\frac{1}{N}.
\end{equation}

The operator $\hat{H}^h_{U}$ is an $h$-differential operator:
\[
\hat{H}^h_{U}=H_0\left(x,\frac{h}{i}\frac{\partial}{\partial x}\right)+hH_1\left(x,\frac{h}{i}\frac{\partial}{\partial x}\right),
\]
where $H_0\in C^\infty(U\times \RR^n)$ is the principal symbol of $\hat{H}^h_{U}$:
\begin{equation}\label{e:defH0}
H_0(x,p) =\sum_{1\leq j,\ell\leq n}g^{j\ell}(x)\left(p_j-A_{U,j}(x)\right)\left(p_\ell-A_{U,\ell}(x)\right)=|p-A_U(x)|^2_{g^{-1}},
\end{equation}
and $H_1\in C^\infty(U\times \RR^n)$ is given by
\begin{equation}\label{e:defH1}
H_1(x,p) =\frac{1}{\sqrt{|g(x)|}}\sum_{1\leq j,\ell\leq n}\frac{1}{i}\frac{\partial}{\partial x^j}\left[\sqrt{|g(x)|}g^{j\ell}(x)\left(p_\ell-A_{U,\ell}(x)\right)\right].
\end{equation}

Let us apply the multi-dimensional WKB-method in the form of the Maslov canonical operator \cite{Maslov65,Maslov-Fed} to the operator $\hat{H}^h_{U}$.

Consider the cotangent bundle $T^*M$ of $M$ with the canonical projection $\pi :T^*M\to M$. Denote by $\Omega_0$ the canonical symplectic form on $T^*M$:
\[
\Omega_0=\sum_{j=1}^ndp_j\wedge dx^j.
\]
Let $\Lambda_0\subset T^*M$ be a Lagrangian submanifold in the phase space $(T^*M,\Omega_0)$. Let us assume that it is oriented and equipped with a smooth volume form $d\mu_0$.  Let $s_0\in \Lambda_0$. Choose a coordinate system $(x_1,\ldots,x_n)$ in a neighborhood $U\subset M$ of $\pi(s_0)$. Let $(x_1,\ldots,x_n, p_1, \ldots, p_n)$ be the corresponding system of canonical coordinates on $\pi^{-1}(U)\subset T^*M$. Write the coordinates of $s\in \Lambda_0$ in the form $(x,p)=(X^{(0)}(s), P^{(0)}(s))$. By the local coordinates lemma, there exists a (possibly, empty) subset $I=\{j_1,\ldots,j_k\}\subset \{1,\ldots,n\}$ such that the function set
\[
(X^{(0)}_I(s),P^{(0)}_{\bar{I}}(s))=(X^{(0)}_{j_1}(s),\ldots, X^{(0)}_{j_k}(s), P^{(0)}_{j_{k+1}}(s), \ldots, P^{(0)}_{j_n}(s)),
\]
where $\bar I=\{j_{k+1},\ldots,j_n\}=\{1,\ldots,n\}\setminus I$, determines a local coordinate system on $\Lambda_0$ in a simply-connected neighborhood $\cU_0\subset \Lambda_0$ of $s_0$. Thus, the equations $x_I=X^{(0)}_I(s)$, $p_{\bar{I}}=P^{(0)}_{\bar{I}}(s)$ are uniquely solvable for $s\in \cU_0$ so that $s=\sigma^{(0)}(x_I, p_{\bar{I}})$. The neighborhood $\cU_0$ endowed with such coordinates $(x_I, p_{\bar{I}})$ is called a canonical chart. We will write the coordinates in $(x_I, p_{\bar{I}})$ in increasing order of indices, that is, as a sequence $(z_1,\ldots,z_n)$, where $z_j=x_j$, if $j\in I$, and $z_j=p_j$, if $j\in \bar{I}$. Similarly, $dx_I\wedge dp_{\bar{I}}=dz_1\wedge \ldots \wedge dz_n$. Then the volume form $d\mu_0$ in the canonical chart $\cU_0$ is written as
\[
d\mu_0=\mathcal J^{(0)}_I dx_I\wedge d p_{\bar{I}},
\]
where $\mathcal J^{(0)}_I(s)$ is the derivative of the measure $d\mu_0$ with respect to the measure $dx_I\wedge d p_{\bar{I}}$ at $s\in \cU$:
\[
\mathcal J^{(0)}_I (s)=\left|\frac{d\mu_0(s)}{dx_I\wedge d p_{\bar{I}}}\right|.
\]

Denote by $\alpha$ the canonical 1-form on $T^*M$:
\[
\alpha=\sum_{j=1}^np_jdx^j.
\]
Since $d\alpha=\Omega_0$ and $\Omega_0\left|_{\Lambda_0}\right.=0$, the form $\alpha\left|_{\Lambda_0}\right.$ is exact in any canonical chart $\cU_0$. Thus, there exists a smooth function $\tau_0$ on $\cU_0$ such that
\[
d\tau_0=\alpha\left|_{\Lambda_0}\right..
\]
The function $\tau_0$ is defined up to an additive constant and given by
\[
\tau_0(s)=\int_{s_0}^s\alpha+\tau_0(s_0), \quad s\in \Lambda_0,
\]
where the integral is taken along any path in $\cU_0$, connecting $s_0$ and $s$. Such a function will be called the eikonal. Fix the eikonal $\tau_0$ and define the action in the canonical chart $\cU_0$ by
\begin{equation}\label{e:SU0}
S_{\cU_0}(s)=\tau_0(s)-\langle P^{(0)}_{\bar I}(s), X^{(0)}_{\bar I}(s)\rangle.
\end{equation}

In the canonical chart $\cU_0$, the local canonical operator
\[
K^h_{\Lambda_0,\cU_0} : C^\infty_0(\cU_0)\to C^\infty(\mathbb R^n)
\]
is defined by the formula (see \cite[\S 8]{Maslov-Fed}, in particular, \cite[(8.25)]{Maslov-Fed})
\begin{multline}\label{e:Kh-Lambda0}
K^h_{\Lambda_0,\cU_0} u(x)= \frac{e^{i\pi |\bar I|/4}}{(2\pi h)^{|\bar I|/2}}\int e^{(i/h)(S_{\cU_0}(\sigma^{(0)}(x_I, p_{\bar{I}}))+\langle p_{\bar{I}}, x_{\bar I}\rangle)}\times \\ \times\sqrt{\mathcal J^{(0)}_I(\sigma^{(0)}(x_I, p_{\bar{I}}))} |g(\pi(x_I, p_{\bar{I}}))|^{-1/4}  u(\sigma^{(0)}(x_I, p_{\bar{I}}))dp_{\bar{I}}
\end{multline}
for any function $u\in C^\infty_0(\cU_0)$. In a non-singular chart ($\bar{I}=\emptyset$) the integral with respect to $p_{\bar{I}}$ is absent, and the formula takes the following form:
\[
K^h_{\Lambda_0,\cU_0} u(x)= e^{(i/h)S_{\cU_0}(\sigma^{(0)}(x))}\sqrt{\mathcal J^{(0)}(\sigma^{(0)}(x))}|g(x)|^{-1/4} u(\sigma^{(0)}(x)).
\]
Taking into account the fact that
\[
\mathcal J^{(0)}(s)|g(\pi(s))|^{-1/2}=\left|\frac{d\mu_0(s)}{dx_g}\right|,
\]
where $dx_g=\sqrt{|g(x)|}dx$ is the Riemannian volume form, this formula can be rewritten as
\begin{equation}\label{e:Kh-Lambda0-non}
K^h_{\Lambda_0,\cU_0} u(x)= e^{(i/h)S_{\cU_0}(\sigma^{(0)}(x))}\sqrt{\left|\frac{d\mu_0(\sigma^{(0)}(x))}{dx_g}\right|} u(\sigma^{(0)}(x)).
\end{equation}

For $x\in\RR^n$, the critical points of the phase are given by $p_{\bar{I}}$ such that the corresponding point $s=\sigma^{(0)}(x_I, p_{\bar{I}})$ belongs to $\Lambda_0$. Therefore, it follows immediately by the stationary phase method that, for any $x\not\in \pi({\rm supp}\,u)$ (in particular, for $x\not\in \pi(\cU_0)$):
\[
K^h_{\Lambda_0,\cU_0} u(x)=O(h^\infty).
\]

The commutation formula holds (see \cite[\S 8]{Maslov-Fed}, in particular, \cite[(8.26)]{Maslov-Fed})
\begin{equation}\label{e:com0}
\hat{H}^h_{U} K^h_{\Lambda_0,\cU_0}u=K^h_{\Lambda_0,\cU_0}\left(\left(P^{(0)}_0-ihP^{(0)}_1\right)u\right)+O(h^{2}),
\end{equation}
for any $u\in C^\infty_0(\cU_0)$, where $P^{(0)}_0$ is the zero order differential operator on $\Lambda_0$ given by
\[
P^{(0)}_0=H_0\left|_{\Lambda_0}\right.,
\]
and, if the manifold $\Lambda_0$ and the form $d\mu_0$ are invariant under the Hamiltonian flow with Hamiltonian $H_0$ with respect to the canonical symplectic form $\Omega_0$ (the invariance of $\Lambda_0$ is equivalent to $H_0\left|_{\Lambda_0}\right.\equiv {\rm const}$), then $P^{(0)}_1$ is the first order differential operator on $\Lambda_0$ given by
\[
P^{(0)}_1=X_{H_0}+\Gamma^{(0)},
\]
where $X_{H_0}$ is the Hamiltonian vector field of the Hamiltonian $H_0$ with respect to the canonical symplectic form $\Omega_0$ and $\Gamma^{(0)}$ is the function on $\Lambda_0$ given by
\[
\Gamma^{(0)}(s)=\sigma_{\rm sub}(\hat{H}^h_{U})(s)-\frac{1}{4}X_{H_0}(\ln |g(\pi(s))|),
\]
where $\sigma_{\rm sub}(\hat{H}^h_{U})$ is the subprincipal symbol of the operator $\hat{H}^h_{U}$:
\begin{equation}\label{e:subprincipal}
\sigma_{\rm sub}(\hat{H}^h_{U})(x,p):=H_1(x,p)-\frac{1}{2} \sum_{j=1}^n\frac{\partial^2H_0}{\partial x_j\partial p_j}(x,p).
\end{equation}
A direct computation shows that
\begin{equation*}
\Gamma^{(0)}(x,p) =\frac{1}{\sqrt{|g(x)|}}\sum_{1\leq j,\ell\leq n}\frac{\partial}{\partial x^j}\left[\sqrt{|g(x)|}\right] g^{j\ell}(x)\left(p_\ell-A_{U,\ell}(x)\right) -\frac{1}{4}X_{H_0}(\ln |g(x)|)=0.
\end{equation*}
Therefore, the operator $P^{(0)}_1$ has the form
\[
P^{(0)}_1=X_{H_0}.
\]

To glue together the local canonical operators constructed in such a way, it is convenient to introduce another symplectic structure in the cotangent bundle, a so called twisted symplectic structure, and, therefore, consider another phase space.

The twisted symplectic form $\Omega$ on $T^*M$ is defined by
\begin{equation}\label{e:twisted}
\Omega=\Omega_0+\pi^* F.
\end{equation}
In local coordinates, this expression is written as
\[
\Omega=\sum_{j=1}^{n}dp_j\wedge dx^j+\sum_{j<k} F_{jk}dx^j\wedge dx^k,
\]
where
\[
F=\sum_{j<k} F_{jk}dx^j\wedge dx^k.
\]

The magnetic geodesic flow $\Phi^t : T^*M\to T^*M$ associated with $(g,F)$ is the Hamiltonian flow with the Hamiltonian
\begin{equation}\label{e:1.2a}
H(x,p)=|p|^2_{g^{-1}}=\sum_{j,k=1}^ng^{jk}p_jp_k,
\end{equation}
with respect to the twisted symplectic form $\Omega$.

Hamilton's equations with an arbitrary Hamiltonian $H$ with respect to $\Omega$ are written as
\begin{equation}\label{e:1.4}
\frac{dx^j}{dt}=\frac{\partial H}{\partial p_j},\quad \frac{dp_j}{dt}=-\frac{\partial H}{\partial x^j}+\frac 12\sum_{k=1}^n F_{jk}\frac{\partial H}{\partial p_k}, \quad j=1,\ldots,n.
\end{equation}
In particular, if $H$ is given by \eqref{e:1.2a}, we obtain the Hamilton system of equations, which determines the magnetic geodesic flow $\Phi^t$:
\begin{equation}\label{e:1.5}
\frac{dx^j}{dt}= 2p^j,\quad \frac{dp_j}{dt}=-\sum_{k,\ell=1}^n\frac{\partial g^{k\ell}}{\partial x^j}p_kp_\ell +\sum_{k=1}^nF_{jk}p^k, \quad j=1,\ldots,n,
\end{equation}
where  $p^j=\sum_{k=1}^ng^{jk}p_k$.

Let us transfer the construction of the local canonical Maslov operator described above to the phase space $(T^*U,\Omega)$.

As above, we assume that  $(x^1,\ldots,x^n)$ are local coordinates on an open subset $U$ of $M$,  the Hermitian line bundle $L$ is trivial on $U$ and $A_U= \sum_{j=1}^nA_{U,j}(x)\,dx^j$ is the corresponding connection form. Consider the map $f_U:T^*U\to T^*U$ given by
\[
f_U(x,p)=(x,p-A_U(x)).
\]
It is easy to check that $f_U^*\Omega=\Omega_0$, that is, $f_U:(T^*U,\Omega_0)\to (T^*U,\Omega)$ is a symplectomorphism. Moreover, the Hamiltonian $f_U^*H$ coincides with the principal symbol $H_0$ of $\hat{H}^h_{U}$. Therefore, $f_U$ takes the Hamilton flow in $(T^*U,\Omega_0)$ with Hamiltonian $H_0$ to the magnetic geodesic flow $\Phi^t$.

Let $\Lambda \subset  T^*U$ be a Lagrangian submanifold in the phase space $(T^*U,\Omega)$. We will assume that it is oriented and equipped with a smooth volume form $d\mu$. Then $\Lambda_0:=f_U^{-1}(\Lambda \cap T^*U)\subset T^*U$ is a Lagrangian submanifold in $(T^*U,\Omega_0)$. We will endow it with the induced orientation and the smooth volume form $d\mu_0=(f_U^{-1})^*d\mu$. If the coordinates of $s\in \Lambda_0$ are given by the functions $x=X^{(0)}(s), p=P^{(0)}(s)$ and the coordinates of $s\in \Lambda$ by the functions $x=X(s), p=P(s)$, then the following relations hold:
\[
X(f_U(s))=X^{(0)}(s),\quad P(f_U(s))=P^{(0)}(s)-A_U(X^{(0)}(s)).
\]

Let $s_0\in \Lambda$ and $s^{(0)}_0=f_U^{-1}(s) \in \Lambda_0$. Consider a canonical chart on $\Lambda_0$, defined in a neighborhood $\cU_0$ of $s^{(0)}_0$. Thus, for $s\in \cU_0 \subset \Lambda_0$, the equations $x_I=X^{(0)}_I(s)$, $p_{\bar{I}}=P^{(0)}_{\bar{I}}(s)$ are uniquely solvable for $s$ so that $s=\sigma^{(0)}(x_I, p_{\bar{I}})$. This chart gives rise to a canonical chart on $\Lambda$, defined in the neighborhood $\cU=f_U(\cU_0)$ of $s_0$. For $s \in \cU$ the equations $x_I=X_I(s)$, $p_{\bar{I}}=P_{\bar{I}}(s)$ are uniquely solvable for $s$ so that $s=\sigma (x_I, p_{\bar{I}})$. Denote by $F_U$ the map $f_U$, considered as a map from $\cU_0$ to $\cU$. The following identities hold:
\begin{equation}\label{e:FU}
F_U(\sigma^{(0)}(x_I, p_{\bar{I}}))=\sigma(x^\prime_I, p^\prime_{\bar{I}}), \quad \sigma^{(0)}(x_I, p_{\bar{I}})=F^{-1}_\cU(\sigma(x^\prime_I, p^\prime_{\bar{I}})),
\end{equation}
where $(x_I, p_{\bar{I}})$ and $(x^\prime_I, p^\prime_{\bar{I}})$ are related by the identities
\begin{equation}
x_I= x^\prime_I,\quad p_{\bar{I}}= \Pi_{\bar{I}}(x^\prime_I, p^\prime_{\bar{I}}):=p^\prime_{\bar{I}}+A_{U,\bar{I}}(X(\sigma(x^\prime_I, p^\prime_{\bar{I}}))). \label{e:FU2}
\end{equation}

The volume form $d\mu$ in the canonical chart $\cU$ is written as
\[
d\mu=\mathcal J_I dx_I\wedge d p_{\bar{I}},
\]
where $\mathcal J_I(s)$ is the derivative of the measure $d\mu$ with respect to the measure $dx_I\wedge d p_{\bar{I}}$ at $s\in \cU$:
\[
\mathcal J_I (s)=\left|\frac{d\mu(s)}{dx_I\wedge d p_{\bar{I}}}\right|.
\]
Since $d\mu=F_U^*d\mu_0$, we have
\begin{equation}\label{e:JJ}
\mathcal J_I(\sigma(x^\prime_I, p^\prime_{\bar{I}}))=\mathcal J^{(0)}_I(\sigma^{(0)}(x^\prime_I, \Pi_{\bar{I}}(x^\prime_I, p^\prime_{\bar{I}})))\left|\frac{\partial \Pi_{\bar{I}}(x^\prime_I, p^\prime_{\bar{I}})}{\partial p^\prime_{\bar{I}}}\right|.
\end{equation}

Consider the corresponding local canonical operator $K^h_{\Lambda_0,\cU_0} : C^\infty_0(\cU_0)\to C^\infty_0(\RR^n)$ defined by \eqref{e:Kh-Lambda0}. There is a well-defined map $F_U^* : C^\infty_0(\cU)\to C^\infty_0(\cU_0)$, induced by the map $F_U$. Put
\[
K^h_{\Lambda,\cU}=K^h_{\Lambda_0,\cU_0}\circ  F_U^* : C^\infty_0(\cU)\to C^\infty(\RR^n).
\]

Assume first that $\cU_0$ is a non-singular chart. Then $\cU$ is a non-singular chart and, by \eqref{e:Kh-Lambda0-non}, we have
\[
K^h_{\Lambda,\cU} u(x)=e^{(i/h)\tau_0(\sigma^{(0)}(x))}\sqrt{\left|\frac{d\mu_0(\sigma^{(0)}(x))}{dx_g}\right|} u(F_U(\sigma^{(0)}(x))).
\]
Define a function $\tau$ on $\cU$ by
\[
\tau(s)=\tau_0(F^{-1}_U(s)),\quad s\in \cU.
\]
Using \eqref{e:FU} and \eqref{e:JJ}, we get
\begin{equation}\label{e:Kh-Lambda-non}
K^h_{\Lambda,\cU} u(x)=e^{(i/h)\tau(\sigma (x))}\sqrt{\left|\frac{d\mu(\sigma(x))}{dx_g}\right|} u(\sigma(x)).
\end{equation}

Let us compute the differential of $\tau$:
\[
d\tau=(F^{-1}_U)^*d\tau_0=(F_U^{-1})^*(\alpha\left|_{\Lambda_0}\right.)=(f_U^{-1})^*\alpha\left|_{\Lambda}\right..
\]
Since
\[
(f_U^{-1})^*\alpha=(f_U^{-1})^*\left(\sum_{k=1}^np_kdx^k\right)=\sum_{k=1}^n(p_k+A_{U,k}(x))dx^k=\alpha+\pi^*A_{U},
\]
we finally get
\[
d\tau=(\alpha+\pi^*A_{U})\left|_{\Lambda}\right..
\]
Therefore, the function $\tau$ is defined up to an additive constant and given by
\begin{equation}\label{e:tau}
\tau(s)=\int_{s_0}^s(\alpha+\pi^*A_{U})+\tau(s_0), \quad s\in \Lambda,
\end{equation}
where the integral is taken along any path in $\cU$, connecting $s_0$ and $s$.

Note that, unlike $\tau_0$, the eikonal $\tau$ is defined locally and depends on the choice of a local trivialization of $L$. If we choose another trivialization of $L$ over $U$ with the corresponding magnetic potential $A^\prime_U$, then
\[
A^\prime_U=A_U+d\psi,
\]
where $e^{i\psi}$ is  the transition function from the trivialization to the another one. By \eqref{e:tau}, we get
\begin{multline*}
\tau^\prime(s)=\tau(s) +\int_{s_0}^s \pi^*d\psi+\tau^\prime(s_0)-\tau(s_0)\\ =\tau(s) +\psi(\pi(s))-\psi(\pi(s_0))+\tau^\prime(s_0)-\tau(s_0), \quad s\in \pi^{-1}(U),
\end{multline*}
therefore, for some $C$, we have
\[
\tau^\prime(s)=\tau(s)+\psi(\pi(s))+C, \quad s\in \pi^{-1}(U).
\]
Putting $C=0$ and $h=1/N$, we obtain that in a non-singular chart the functions
\[
E(x)=e^{(i/h)\tau(\sigma(x))}, \quad E^\prime(x)=e^{(i/h)\tau^\prime(\sigma(x))}, \quad x\in U,
\]
are related by the identity
\[
E^\prime(x)=e^{iN\psi(x)}E(x), \quad x\in U.
\]
This shows that the functions $K^h_{\Lambda,\cU} u(x)$, defined by \eqref{e:Kh-Lambda-non} for different trivializations of $L$, give rise to a section of the bundle  $L^N$.

Consider the case of an arbitrary canonical chart $\cU$. By \eqref{e:Kh-Lambda0}, we get
\begin{multline*}
K^h_{\Lambda,\cU} u(x)= \frac{e^{i\pi |\bar I|/4}}{(2\pi h)^{|\bar I|/2}}\int e^{(i/h)(S_{\cU_0}(\sigma^{(0)}(x_I, p_{\bar{I}}))+\langle p_{\bar{I}}, x_{\bar I}\rangle)}\times \\ \times \sqrt{\mathcal J^{(0)}_I(\sigma^{(0)}(x_I, p_{\bar{I}}))}|g(\pi(\sigma^{(0)}(x_I, p_{\bar{I}})))|^{-1/4} u(F_U(\sigma^{(0)}(x_I, p_{\bar{I}}))) dp_{\bar{I}}.
\end{multline*}
Let us make the change of variables in the integral in the right-hand side of this formula given  by \eqref{e:FU2} and use the relations \eqref{e:FU}:
\begin{multline*}
K^h_{\Lambda,\cU} u(x)= \frac{e^{i\pi |\bar I|/4}}{(2\pi h)^{|\bar I|/2}}\int e^{(i/h)(S_{\cU_0}(F_U^{-1}(\sigma(x^\prime_I, p^\prime_{\bar{I}})))+\langle \Pi_{\bar{I}}(x^\prime_I, p^\prime_{\bar{I}}), x^\prime_{\bar I}\rangle)}\times \\ \times \sqrt{\mathcal J^{(0)}_I(F_U^{-1}(\sigma(x^\prime_I, p^\prime_{\bar{I}})))}|g(\pi(\sigma(x^\prime_I, p^\prime_{\bar{I}})))|^{-1/4}\left|\frac{\partial \Pi_{\bar{I}}(x^\prime_I, p^\prime_{\bar{I}})}{\partial p^\prime_{\bar{I}}}\right|  u(\sigma(x^\prime_I, p^\prime_{\bar{I}}))  dp^\prime_{\bar{I}}.
\end{multline*}
Define the action $S_{\cU}$ on $\cU$ by
\[
S_{\cU}(\sigma(x^\prime_I, p^\prime_{\bar{I}})):=S_{\cU_0}(F_U^{-1}(\sigma(x^\prime_I, p^\prime_{\bar{I}}))).
\]
According to \eqref{e:SU0}, we get
\begin{align*}
S_{\cU}(\sigma(x^\prime_I, p^\prime_{\bar{I}})) & =\tau_0(F_U^{-1}(\sigma(x^\prime_I, p^\prime_{\bar{I}})))-\langle P^{(0)}_{\bar I}(F_U^{-1}(\sigma(x^\prime_I, p^\prime_{\bar{I}}))), X^{(0)}_{\bar I}(F_U^{-1}(\sigma(x^\prime_I, p^\prime_{\bar{I}})))\rangle\\ & =\tau(\sigma(x^\prime_I, p^\prime_{\bar{I}}))-\langle \Pi_{\bar{I}}(x^\prime_I, p^\prime_{\bar{I}}), X_{\bar I}(\sigma(x^\prime_I, p^\prime_{\bar{I}}))\rangle.
\end{align*}
By \eqref{e:JJ}, we have
\[
\mathcal J^{(0)}_I(F_U^{-1}(\sigma(x^\prime_I, p^\prime_{\bar{I}})))=\mathcal J^{(0)}_I(\sigma^{(0)}(x^\prime_I, \Pi_{\bar{I}}(x^\prime_I, p^\prime_{\bar{I}})))=\mathcal J_I(\sigma(x^\prime_I, p^\prime_{\bar{I}}))\left|\frac{\partial \Pi_{\bar{I}}(x^\prime_I, p^\prime_{\bar{I}})}{\partial p^\prime_{\bar{I}}}\right|^{-1}.
\]
Using these formulas, we arrive at the following formula for the canonical operator in the chart  $\cU$:
\begin{multline}\label{e:Kh-Lambda-2}
K^h_{\Lambda,\cU} u(x)= \frac{e^{i\pi |\bar I|/4}}{(2\pi h)^{|\bar I|/2}}\int e^{(i/h)(S_{\cU}(\sigma(x^\prime_I, p^\prime_{\bar{I}}))+\langle \Pi_{\bar{I}}(x^\prime_I, p^\prime_{\bar{I}}), x_{\bar I}\rangle)}\times \\ \times \sqrt{\mathcal J_I(\sigma(x^\prime_I, p^\prime_{\bar{I}}))}|g(\pi(\sigma(x^\prime_I, p^\prime_{\bar{I}})))|^{-1/4} \left|\frac{\partial \Pi_{\bar{I}}(x^\prime_I, p^\prime_{\bar{I}})}{\partial p^\prime_{\bar{I}}}\right|^{1/2}u(\sigma(x^\prime_I, p^\prime_{\bar{I}})) dp^\prime_{\bar{I}}.
\end{multline}

Let us rewrite the commutation formula \eqref{e:com0}. For $u\in C^\infty_0(\cU)$, we get
\begin{align*}
\hat{H}^h_{U}K^h_{\Lambda,\cU}u &= \hat{H}K^h_{\Lambda_0,\cU_0}\circ  F_U^*u \\ & =K^h_{\Lambda_0,\cU_0}\left(\left(P^{(0)}_0-ihP^{(0)}_1\right)(F_U^*u)\right)+O(h^2) \\ &
=K^h_{\Lambda,\cU}\left(\left((F_U^{-1})^*P^{(0)}_0F^*_U-ih(F_U^{-1})^*P^{(0)}_1F^*_U\right)u\right)+O(h^2)\\ &=
K^h_{\Lambda,\cU}\left(\left(P_0-ihP_1\right)u\right)+O(h^2),
\end{align*}
where $P_0$ is the zero order differential operator on $\Lambda$, given by
\begin{equation}\label{e:def-P0}
P_0=(F_U^{-1})^*P^{(0)}_0F^*_U =H\left|_{\Lambda}\right.,
\end{equation}
and if the manifold $\Lambda$ and the form $d\mu$ are invariant under the magnetic geodesic flow (the invariance of $\Lambda$ is equivalent to $H\left|_{\Lambda}\right.\equiv {\rm const}$), then $P_1$ is the first order differential operator on $\Lambda$ given by
\begin{equation}\label{e:def-P1}
P_1=(F_U^{-1})^*P^{(0)}_1F^*_U=X_{H},
\end{equation}
where $X_{H}$ is the Hamiltonian vector field with the Hamiltonian $H$ with respect to the twisted symplectic form $\Omega$ (the differentiation operator along the trajectories of the magnetic geodesic flow).

The next step of the construction consists in constructing the global canonical operator. First, we have to fix a Lagrangian submanifold $\Lambda$ in the phase space $(T^*M,\Omega)$.  The previous construction allows us to construct a local canonical operator in each canonical chart on $\Lambda$. To glue together such local operators, we have to compare local canonical operators constructed in two canonical charts on the intersection of charts and try to get that they coincide. It is well-known that this cannot be done for an arbitrary Lagrangian submanifold $\Lambda$. Certain conditions, called quantization conditions, should be satisfied. If $\Lambda$ satisfies these conditions, then we get the global canonical operator $K_\Lambda^{1/N} : C^\infty_0(\Lambda) \to C^\infty(M,L^N)$. The following commutation formula holds:
\[
\Delta^{L^N}K_\Lambda^{1/N}u=K_\Lambda^{1/N}\left(\left(N^2P_0-iNP_1\right)u\right)+O(1),
\]
for any $u\in C^\infty_0(\Lambda)$, where the differential operators $P_0$ and $P_1$ on $\Lambda$ are given by \eqref{e:def-P0} and \eqref{e:def-P1}, respectively.

Using these facts, one can construct almost eigenfunctions for the operator $\Delta^{L^N}$. More precisely, suppose that $H\left|_\Lambda\right.\equiv E$ and $u\in C^\infty_0(\Lambda)$ be a function such that
\begin{equation}\label{e:P1=0}
P_1u=\kappa u.
\end{equation}
Then the following relation holds:
\[
\Delta^{L^N}K_\Lambda^{1/N}u=K_\Lambda^{1/N}\left(N^2P_0u-iNP_1u\right)+O(1)=(E N^2 -i\kappa N) K_\Lambda^{1/N}u +O(1).
\]
Thus, the section $U_N=K_\Lambda^{1/N}u \in C^\infty(M,L^N)$ is an almost eigenfunction  of $\Delta^{L^N}$ with the corresponding eigenvalue $\hat\lambda_N=E N^2-i\kappa N$.

Admissible values of $E$ are found from the quantization condition on $\Lambda$.  First, recall the notion of action for closed curves in $T^*M$. For a closed curve $\gamma$ in $T^*M$, denote by $\exp (ih_A(\gamma))\in U(1)$ the holonomy of its projection $\pi\circ \gamma$ to $M$ with respect to the connection $\nabla^L$ on $L$. Then the action $S_\gamma$ of $\gamma$ is defined modulo multiplies of $2\pi$ by
\[
S_\gamma=\int_\gamma \sum_{j=1}^np_jdx^j+ h_A(\gamma) \mod 2\pi\ZZ.
\]
The quantization condition on the Lagrangian submanifold $\Lambda$ reads as follows. For any closed curve $\gamma$ on $\Lambda$, we have
\[
NS_\gamma =\frac{\pi}{2}l_\gamma \mod 2\pi\ZZ,
\]
where $l_\gamma \in \ZZ$ is the Maslov index of $\gamma$.

As a solution of \eqref{e:P1=0} with $\kappa=0$, one can take the function
\begin{equation}\label{e:P1=0-solution}
u(x)\equiv 1.
\end{equation}

We will not discuss these questions in the general case, and consider the simplest example of magnetic monopole --- the case of the Dirac magnetic monopole on the two-dimensional sphere.

\section{Dirac magnetic monopole}
Let the Riemannian manifold $(M,g)$ be the two-dimensional sphere $S^2$ in $\RR^3$ of radius 1 with center at the origin:
\[
S^2=\{(x,y,z)\in \RR^3 : x^2+y^2+z^2=1\},
\]
equipped with the Riemannian metric induced by the embedding in the Euclidean space $\RR^3$. In the spherical coordinates
\[
x=\sin\theta \cos\varphi, \quad y=\sin\theta \sin\varphi, \quad z=\cos\theta, \quad \theta\in (0,\pi), \varphi\in (0,2\pi),
\]
the Riemannian metric $g$ is given by
\[
g=d\theta^2+\sin^2\theta d\varphi^2.
\]
Consider a magnetic field form $F$ of the form
\[
F=B\,d \mathrm{vol}_M=B\sin\theta d\theta\wedge d\varphi.
\]
The quantization condition \eqref{e:prequantum} means that
\begin{equation}\label{e:sphere-quant}
B\in \frac 12\ZZ.
\end{equation}

For $B=N/2, N\in \ZZ,$ the corresponding Hermitian line bundle $L^N$ can be described as the line bundle associated with the Hopf fibration $S^3\to S^2$ and the character $\chi_N : S^1\to S^1$ given by $ \chi_N(u)=u^N, u\in S^1$. As already mentioned in Introduction, in physical literature, this quantum system is a well-known Wu-Yang magnetic monopole \cite{WY}, which provides a natural topological interpretation of the Dirac monopole with magnetic charge  $\mu=N\hbar c/(2e)$.

Consider a cover of the sphere by two coordinate neighborhoods $U_1$ and $U_2$:
\[
U_1=\left\{0 \leq \theta <\pi, 0\leq \varphi <2\pi\right\}, \quad
U_2=\left\{0 < \theta \leq \pi, 0\leq \varphi <2\pi\right\}.
\]
Sections of $L^N$ are given by sets of functions $\xi_1$ on $U_1$ and $\xi_2$ on $U_2$ such that, on $U_1\cap U_2$, we have
\begin{equation}\label{e:transition}
\xi_1=g_{12}\xi_2,
\end{equation}
where $g_{12}$ is the transition function:
\[
g_{12}=e^{2iB\varphi}.
\]
The magnetic potentials are given by:
\[
A_1=B(1-\cos\theta)d\varphi, \quad \text{on}\ U_1;\quad
A_2=-B(1+\cos\theta)d\varphi, \quad \text{on}\ U_2.
\]
On $U_1\cap U_2$, they are related by the identity
\[
A_1-A_2=g^{-1}_{12}dg_{12}.
\]
The magnetic Laplacian $\Delta^{L^N}$ is given by:
\[
\Delta^{L^N}=-\frac{1}{\sin\theta}\frac{\partial}{\partial\theta}\sin\theta\frac{\partial}{\partial\theta}-\frac{1}{\sin^2\theta}\left(\frac{\partial}{\partial\varphi}-iB(1-\cos\theta)\right)^2 \quad \text{on}\ U_1;
\]
\[
\Delta^{L^N}=-\frac{1}{\sin\theta}\frac{\partial}{\partial\theta}\sin\theta\frac{\partial}{\partial\theta}-\frac{1}{\sin^2\theta}\left(\frac{\partial}{\partial\varphi}+iB(1+\cos\theta)\right)^2\quad \text{on}\ U_2.
\]

The spectrum of the magnetic Laplacian $\Delta^{L^N}$ is computed in
 \cite{Tamm,WY}. It consists of the eigenvalues
\[
\lambda_{N,j}=j(j+1)+\frac{N}{2}(2j+1), \quad j=0,1, \ldots,
\]
with multiplicity
\[
m_{N,j}=N+2j+1.
\]
the corresponding eigenfunctions are known as monopole harmonics.

The magnetic geodesic flow is given by the Hamiltonian
\[
H(\theta, \varphi, p_\theta, p_\varphi)=p_\theta^2+\frac{1}{\sin^2\theta}p^2_\varphi
\]
with respect to the twisted symplectic form $\Omega$ on $T^*M$:
\[
\Omega=dp_\theta\wedge d\theta+dp_\varphi\wedge d\varphi +B\sin\theta d\theta\wedge d\varphi.
\]
The corresponding Hamilton equations have the form:
\[
\dot{\theta}=2p_\theta, \quad \dot{\varphi}=\frac{2}{\sin^2\theta}p_\varphi, \quad \dot{p}_\theta=-\frac{2\cos\theta}{\sin^3\theta}p^2_\varphi+B\sin\theta p_\varphi, \quad \dot{p}_\varphi=-B\sin\theta p_\theta.
\]

As already said in Introduction, in this case, the magnetic geodesic flow is superintegrable. Therefore, the division of the phase space into invariant tori is not unique and determined, for instance, by the choice of two commuting first integrals $I_1, I_2$. As one of them, it is natural to take the Hamiltonian:
\[
I_1=H(\theta, \varphi, p_\theta, p_\varphi).
\]
As an additional first integral, we take
\[
I_2=p_\varphi-B\cos\theta.
\]
It is easy to check that $I_2$ is everywhere defined.

The corresponding invariant tori $\Lambda$ are parameterized by two parameters $E$ and $P$ and given by the equations
\[
p_\theta^2+\frac{1}{\sin^2\theta}p^2_\varphi=E, \quad p_\varphi-B\cos\theta=P.
\]
Thus, for any $(\theta,\varphi, p_\theta, p_\varphi)\in \Lambda=\Lambda (E,P)$, the following formulas hold:
\begin{equation}\label{e:ptheta-pvarphi}
p_\theta=\pm\left(E-\frac{1}{\sin^2\theta}(P+B\cos\theta)^2\right)^{1/2}, \quad p_\varphi=P+B\cos\theta.
\end{equation}
The torus $\Lambda (E,P)$ is non-empty if and only if the following inequality has a solution:
\[
E-\frac{1}{\sin^2\theta}(P+B\cos\theta)^2\geq  0,
\]
or equivalently
\[
E-P^2-2BP\cos\theta-(E+B^2)\cos^2\theta \geq  0.
\]
Therefore, if we consider the quadratic function $R(z)=a_1+b_1z+c_1z^2$ with the coefficients $a_1=E-P^2$, $b_1=-2BP$, $c_1=-(E+B^2)$, then the condition should hold:
\[
\Delta_1:=b_1^2-4a_1c_1=4(E^2+E(B^2-P^2))>0,
\]
which implies the following condition on $E$ and $P$:
\[
P^2<E+B^2.
\]
The roots of $R(z)$ are given by
\[
z_1=\frac{-BP-\sqrt{E^2+E(B^2-P^2)}}{E+B^2}, \quad
z_2=\frac{-BP+\sqrt{E^2+E(B^2-P^2)}}{E+B^2}.
\]
One can check that $z_2\leq 1$, moreover $z_2=1 \Leftrightarrow P=-B$. Similarly, $z_1\geq -1$, moreover $z_1=-1 \Leftrightarrow P=B$.

Thus, for any $(\theta,\varphi, p_\theta, p_\varphi)\in \Lambda$, we have the relation
\[
\cos \theta\in [z_1, z_2]\Leftrightarrow \theta\in [\theta_2, \theta_1], \quad \theta_j:=\arccos z_j, j=1,2,
\]
which describes the image of $\Lambda$ under the canonical projection $\pi :T^*S^2\to S^2$:
\[
\pi(\Lambda)=\{(\theta,\varphi) : \theta_2 \leq \theta \leq \theta_1, 0\leq \varphi <2\pi\}.
\]
Singular points of the restiction of $\pi$ to $\Lambda$ occur when $\theta=\theta_1$ or $\theta=\theta_2$. Therefore, the singularity cycle $\Sigma(\Lambda)$ consists of two circles
\[
\{ p_\theta=0, p_\varphi=P+B\cos\theta_j, \theta=\theta_j, \varphi\in [0,2\pi)\}, \quad j=1,2.
\]

One can introduce two non-singular canonical charts $\cU^\pm$ on the torus $\Lambda$  with coordinates
\[
(\theta,\varphi)\in U:=\left\{\theta_2 < \theta <\theta_1, 0\leq \varphi <2\pi\right\} \mapsto (\theta,\varphi, p_\theta, p_\varphi)\in \cU^\pm\subset \Lambda,
\]
where $p_\theta$ and $p_\varphi$ are determined by \eqref{e:ptheta-pvarphi}.

Here we meet the simplest case when the projection $\pi(\cU^\pm)$ of the canonical chart $\cU^\pm$ to $M$ is contained in two different coordinate neighborhoods $U_1$ and $U_2$. In spite of the fact that the local coordinates on $\pi(\cU^\pm)$ defined by the local coordinates on $U_1$ and $U_2$ coincide, trivializations of $L$ over them are different, that leads to  different values of the eikonal functions $\tau^\pm_{U_1}$ and $\tau^\pm_{U_2}$.

If $\pi(\cU^\pm)$ is considered as a subset of $U_1$, then
\begin{multline*}
d\tau^\pm_{U_1}=p_\theta d\theta+(p_\varphi+B(1-\cos\theta))d\varphi \\
=\pm\left(E-\frac{1}{\sin^2\theta}(P+B\cos\theta)^2\right)^{1/2}d\theta+(P+B)d\varphi, \quad (\theta,\varphi)\in U.
\end{multline*}
If $\pi(\cU^\pm)$ is considered as a subset of $U_2$, then
\begin{multline*}
d\tau^\pm_{U_2}=p_\theta d\theta+(p_\varphi-B(1+\cos\theta))d\varphi\\ =\pm\left(E-\frac{1}{\sin^2\theta}(P+B\cos\theta)^2\right)^{1/2} d\theta+(P-B)d\varphi,  \quad (\theta,\varphi)\in U.
\end{multline*}
From here, we find:
\[
\tau^\pm_{U_1}(\theta,\varphi)=\pm I(\theta)+(P+B)\varphi+\tau_1,\quad
\tau^\pm_{U_2}(\theta,\varphi)=\pm I(\theta)+(P-B)\varphi+\tau_2,
\]
where
\begin{multline*}
I(\theta)=\int \left(E-\frac{1}{\sin^2\theta}(P+B\cos\theta)^2\right)^{1/2}d\theta \\
\begin{aligned}
= &\frac 12 |P+B|\arcsin \frac{2a_1+b_1+(b_1+2c_1)\cos\theta}{(\cos\theta-1)\sqrt{\Delta_1}}\\
& +\frac 12 |P-B|\arcsin \frac{2a_1-b_1+(b_1-2c_1)\cos\theta}{(\cos\theta+1)\sqrt{\Delta_1}}\\
& +\sqrt{E+B^2}\arcsin \frac{2c_1\cos\theta+b_1}{\sqrt{\Delta_1}}.
\end{aligned}
\end{multline*}
$\tau_1$, $\tau_2$ are some constants.

It is easy to see that, for $\tau_1=\tau_2$, the functions $\exp\left( iN\tau^\pm_{U_j}\right)$ on $\cU^\pm$ satisfy the compatibility condition \eqref{e:transition}.

Put $B=1/2$. The $2\pi$-periodicity condition for the function $\exp\left( iN\tau^\pm_{U_j}\right)$ is valid, if, for some $k_1\in \ZZ$,
\[
\int_0^{2\pi}\left(P+\frac 12\right)d\varphi=2\pi\left(P+\frac 12\right)=2\pi \frac{k_1}{N}.
\]
Here one should note that, since the magnetic field quantization condition \eqref{e:sphere-quant} is satisfied, the $2\pi$-periodicity of the function $\exp\left( iN\tau^\pm_{U_1}\right)$ in $\varphi$ immediately implies the $2\pi$-periodicity of the function $\exp\left( iN\tau^\pm_{U_2}\right)$ in $\varphi$. If \eqref{e:sphere-quant} is not satisfied (that is, $B\not\in \frac 12\ZZ$), then the functions $\exp\left( iN\tau^\pm_{U_1}\right)$ and $\exp\left( iN\tau^\pm_{U_2}\right)$ can not simultaneously be $2\pi$-periodic for any value of $P$.

The complete integral
\[
\mathcal J=\int_\gamma \left(E-\frac{1}{\sin^2\theta}(P+B\cos\theta)^2\right)^{1/2}d\theta
\]
along the closed cycle $\gamma =\{\theta\in [\theta_2, \theta_1], \varphi\in [0,2\pi]\}$ on $\Lambda$ equals
\[
\mathcal J=2(I(\theta_1)-I(\theta_2))= -\pi |P+B|-\pi |P-B|+2\pi\sqrt{E+B^2}.
\]
The fact that the functions $\exp\left( iN\tau^\pm_{U_j}\right)$ in $\theta$ are well-defined (taking into account the Maslov index, that is, a condition $\tau_1-\tau_2$) is true, if, for some $k_2\in \ZZ$
\[
2\pi\sqrt{E+\frac 14} -\pi \left|P+\frac 12\right|-\pi \left|P-\frac 12\right|=2\pi \frac{k_2+\frac 12}{N}.
\]
Taking into account that $P=\frac{k_1}{N}-\frac 12$, we get:
\[
\sqrt{E+\frac 14} = \frac{j+\frac 12}{N}+\frac 12,
\]
where
\[
j=\begin{cases}
-k_1+k_2, & \text{if}\ k_1<0, \\
k_2, & \text{if}\ 0\leq k_1<N, \\
k_1-N+k_2, & \text{if}\ k_1\geq N.
\end{cases}
\]
Since $P^2<E+B^2$, the relation $k_2\geq 0$ holds and, therefore, $j\geq 0$.

Thus, admissible values of the parameter $E$, given by the quantization condition, have the form
\[
E_{N,j}=\frac{j(j+1)+\frac{N}{2}(2j+1)+\frac 14}{N^2},\quad j=0,1,2,\ldots.
\]
As a solution to \eqref{e:P1=0} with $\kappa=0$, we take the function given by  \eqref{e:P1=0-solution}:
\[
u(\theta,\varphi)\equiv 1.
\]
By \eqref{Delta-hatH}, we get the following formula for the almost eigenvalues $\hat\lambda_{N,j}$ of $\Delta^{L^N}$:
\begin{equation}\label{e:hat-lambda}
\hat\lambda_{N,j}=E_{N,j}N^{2}=j(j+1)+\frac{N}{2}(2j+1)+\frac 14, \quad j=0,1, \ldots.
\end{equation}

For fixed $N$ and $j$,  admissible values of the parameter $P$ are given by the conditions
\[
k_1=N\left(P+\frac 12\right)\in \ZZ, \quad -j \leq k_1\leq N+j.
\]
Therefore, the multiplicity of $\hat\lambda_{N,j}$ is equal to
\begin{equation}\label{e:m-lambda}
\hat m_{N,j}=N+2j+1.
\end{equation}

Recall the formulas for exact eigenvalues:
\[
\lambda_{N,j}=j(j+1)+\frac{N}{2}(2j+1), \quad j=0,1, \ldots,
\]
and their multiplicities
\[
m_{N,j}=N+2j+1.
\]
We see that  the formula \eqref{e:hat-lambda} describes exact eigenvalues of the magnetic Laplacian up to a constant correction term $\Delta\lambda_{N,j}=\frac 14$, and the formula  \eqref{e:m-lambda} gives a correct answer for the multiplicities of these eigenvalues.

\end{document}